\documentclass[onecolumn]{emulateapj}
\usepackage{longtable}
\usepackage{graphicx}
\usepackage{graphics}
\usepackage{color}
\usepackage{epsfig}
\usepackage{rotating}
\usepackage{subfigure}
\usepackage{float}

\shortauthors{}
\shorttitle{}
\begin{document}

\title{\emph{Chandra} Survey of Nearby Highly Inclined Disc Galaxies - II:\\Correlation Analysis of Galactic Coronal Properties}

\author{Jiang-Tao Li\altaffilmark{1,2} and Q. Daniel Wang\altaffilmark{1}} \altaffiltext{1}{Department of Astronomy, University of
Massachusetts, 710 North Pleasant Street, Amherst, MA 01003, U.S.A.}\altaffiltext{2}{Service d'Astrophysique (SAp)/IRFU/DSM/CEA Saclay, Bt. 709, 91191 Gif-sur-Yvette Cedex, France}

\keywords{galaxies: general---galaxies: halos---galaxies: normal---X-rays: galaxies}

\nonumber

\begin{abstract}
X-ray observations provide a key tool for exploring the properties of galactic coronae and their formation processes. In an earlier paper, we have presented a \emph{Chandra} data analysis of the coronae of 53 nearby highly-inclined disc galaxies. Here we study the correlation of the X-ray measurements of the coronae with other galaxy properties and compare the results with those obtained for elliptical galaxies. A good correlation is present between the coronal luminosity ($L_X$) and the star formation rate (SFR). But we find a better correlation between $L_X$ and the total SN mechanical energy input rate ($\dot{E}_{SN}$), including the expected contribution from both core collapsed (CC) and Type~Ia SNe. The X-ray radiation efficiency ($\eta\equiv L_{X}/\dot{E}_{SN}$) of the coronae has a mean value of $\sim0.4\%$ with an $rms$ of $0.50\pm0.06\rm~dex$. $\eta$ further correlates with $M_{TF}/M_*$ ($M_{TF}$ is the total baryon mass measured from the rotation velocity and the Tully-Fisher relation, $M_*$ is the stellar mass measured from the K-band luminosity) and the CC SN rate surface density ($F_{SN(CC)}$, in units of $\rm SN~yr^{-1}~kpc^{-2}$), which can be characterized as: $\eta=(0.41_{-0.12}^{+0.13}\%) M_{TF}/M_*$ and $\eta=(1.4\pm0.5\%)F_{SN(CC)}^{-(0.29\pm0.11)}$. These correlations reflect the roles that played by the gravitational mass and energetic feedback concentrations of the galaxies in determining their X-ray radiation efficiency. The characteristic temperature ($T_X$) of the coronal gas shows little dependence on the total or specific SFR, the cold gas content, or $L_X$. The coronae of disc galaxies tend to be more X-ray luminous, hotter, and lower in the Fe/O abundance ratio than those of elliptical ones of similar masses. Early-type non-starburst disc galaxies tend to be more Fe-rich, while starburst ones have a roughly constant abundance ratio of ${\rm Fe/O}\sim0.36\pm0.12\rm~solar$. Our results are consistent with the coronal gas being mainly provided by stellar feedback in a galaxy stellar mass range of $\sim10^{8.7-11}\rm~M_\odot$. In addition, processes such as charge exchange at cool/hot gas interfaces, as well as various other environmental effects, are also needed to explain the diversity of the observed coronal X-ray properties.
\end{abstract}

\section{Introduction}\label{PaperIIsec:Introduction}

Galactic coronae around disc galaxies play a key role in galaxy formation and evolution. Such coronae are believed to serve not only as reservoirs from which galaxies acquire baryons to build up and/or sustain galactic discs \citep{White91,Benson00,Toft02,SommerLarsen06,Rasmussen09,Bogdan12}, but also as depositories of galactic feedback from various stellar descendants and supermassive black holes (BHs) \citep{Strickland00a,Strickland02,Strickland04a,Strickland04b,Tullmann06a,Tullmann06b,Li08}. Study of the coronae thus helps us to understand how mass, energy, and heavy elements circulate in various galaxy environments (e.g., \citealt{Wang10,Anderson12}).

Two major scenarios have been proposed for the origin of the coronae around disc galaxies: One links the diffuse X-ray emission in and around these galaxies to the feedback processes of active galactic nuclei (AGN) \citep{Kraft05,Forman05,McNamara12}, star formation (SF) \citep{Strickland04a,Strickland04b,Grimes05,Li08}, and/or mass-loss and Type~Ia SNe of evolved low-mass stars \citep{Li09,Li11}; The other associates the presence of large-scale diffuse hot gas with the accretion of the intergalactic medium (IGM) (e.g., \citealt{White91}). In this latter scenario, the density of the accretion-shock-heated hot gas increases toward the optical disk of a massive galaxy, resulting in enhanced X-ray emission that may be detectable (e.g., \citealt{Benson00,Toft02,Anderson11,Anderson12,Bogdan12,Dai12}).

The above two scenarios are not unambiguously distinguishable, especially with little observational constraint on the hot gas bulk motion \citep{Crain10a}. In current galaxy formation models, the growth of stellar mass (i.e., the SFR) is closely related to the gravity or the mass of the dark matter halo which also determines the radiative cooling rate (i.e., the X-ray emission) of the accreted gas \citep{Dave08,Dave11}. Therefore, the coronal soft X-ray luminosity, the galaxy stellar mass, the SFR, and the host halo mass are expected to correlate with each other, no matter how the coronae are heated. Both cosmological hydrodynamical simulations (e.g., \citealt{Crain10a}) and observational studies with limited sample sizes ($\lesssim 10$ galaxies, e.g., \citealt{Strickland04b,Grimes05,Tullmann06b}) have reproduced such correlations. However, the observed correlations, or the scaling relations of the coronal luminosity with the stellar mass (traced by the near-IR or optical luminosity) and SFR (often traced by far-IR or radio luminosity), alone are of little use in distinguishing the feedback and accretion scenarios. In addition, one also needs to treat self-consistently the feedback from both young and old stellar populations; their relative importance changes with galaxy types. These issues cannot be addressed with small samples of disc galaxies.

In addition to the above two major scenarios in producing galactic coronae, there are some other mechanisms which could regulate, or even sometimes play a dominant role in determining, the X-ray properties of a corona. For example, it has been argued that a large fraction of the soft X-ray emission from nearby galaxies may actually arise directly from hot and cool gas interfaces \citep{Li09,Li11}, including shock-heating, turbulent mixing (e.g., \citealt{Melioli04}), and charge exchange (CXE) \citep{Liu10,Liu11,Liu12}. Such processes may significantly enhance the soft X-ray emission, as well as lead to underestimation of hot gas temperature. Another possible mechanism strongly affecting the coronal properties is the environment of a galaxy (e.g., \citealt{Mulchaey10}). For example, the luminosity of a corona can be enhanced due to the compression/confinement by the thermal/ram pressure of the intra-cluster medium (ICM) \citep{Lu11}, by the presence of cool gas surrounding a galaxy (e.g., \citealt{Lehnert99}), or by the accretion of external gas \citep{Bogdan11}. On the other hand, strong ram-pressure stripping may sometimes significantly reduce the luminosity of a corona \citep{Bahe12a,Bahe12b}. These secondary effects, which are poorly understood theoretically, cannot be well characterized observationally with small samples of galaxies selected only for certain types or certain environments.

In Paper~I \citep{Li13a}, we have built up a database of galactic coronae in and around 53 nearby highly-inclined disc galaxies, over broad ranges of SFR, morphological type, stellar mass, and clustering environment (see \S\ref{PaperIIsubsec:paperI} for a brief review). We have provided systematical measurements of multiple coronal parameters, such as the diffuse soft X-ray scale height ($h_{exp}$), hot gas luminosity ($L_X$), temperature ($T_X$), and iron-to-oxygen abundance ratio (Fe/O), as well as many derived parameters, such as density ($n_e$), mass ($M_{hot}$), thermal energy ($E_{hot}$), and radiative cooling timescale ($t_{cool}$). Compared to previous works, we benefit from a thorough subtraction of X-ray bright sources in the disc and a uniform data reduction of a large sample of galaxies with various properties. In the present paper, we study the correlation of these hot gas parameters with other galaxy parameters: the SFR, the infrared (IR) warmth ($f_{60}/f_{100}$, where $f_{60}$ and $f_{100}$ are the \emph{IRAS} 60 and 100~$\rm\mu m$ fluxes), the stellar mass ($M_*$), the mass derived from the rotation velocity ($v_{rot}$) and the baryonic Tully-Fisher relation ($M_{TF}$), the atomic ($M_{HI}$) and molecular ($M_{H_2}$) gas masses, the environmental density ($\rho$), and the morphological type code (TC; by definition, the larger the TC value, the later-type a galaxy, with TC=0 roughly separating spiral and S0 galaxies). In particular, we will study the secondary correlations (after taking out the first-order scaling relations of $L_X$ with SFR and/or $M_*$) in various subsamples. Comparisons with recent cosmological simulations will be presented in a follow-up paper (Paper~III, \citealt{Li13c}).

The present paper is organized as follows: In \S\ref{PaperIIsec:SampleDescription}, we summarize the key relevant information about the galaxy sample and the X-ray measurements (as detailed in Paper~I) and further define several physically meaningful parameters from the combination of the directly measured ones. We present the results from correlation analysis of the coronal and other galaxy properties in \S\ref{PaperIIsec:correlation}, in particular, searching for the significant secondary correlations. In \S\ref{PaperIIsec:Discussion}, we compare our measurements with those of elliptical galaxies, and further explore the implications of the results. The main results and conclusions are summarized in \S\ref{PaperIIsec:Summary}.

\section{Description of Galaxy Sample and Parameters}\label{PaperIIsec:SampleDescription}

\subsection{Summary of Paper~I}\label{PaperIIsubsec:paperI}

Here we briefly describe the sample selection, multi-wavelength parameter estimation, subsample definition, X-ray data reduction/measurement, and various potential complications in the characterization of the galactic coronae (see Paper I for details).

The sample is selected with the following criteria: (1) optical morphological type code $-3\lesssim TC\lesssim9$ (ranging from lenticular to bulgeless spiral); (2) inclination angle $i\gtrsim60^\circ$; (3) distance $d\lesssim30\rm~Mpc$; (4) foreground absorption column density $N_H\lesssim8\times10^{20}\rm~cm^{-2}$; (5) optical diameter $1^\prime\lesssim D_{25}\lesssim16^\prime$; (6) total non-grating \emph{Chandra}/ACIS exposure $t_{exp}\gtrsim10\rm~ks$, with two exceptions, NGC~660 ($\sim7.1\rm~ks$) and NGC~4666 ($\sim5\rm~ks$); (7) no bright AGN which would prevent an effective study of the galactic corona because of scattered X-ray photons. Multi-wavelength galaxy properties are calculated and/or listed, including SFR from the \emph{IRAS} IR luminosity $L_{IR}$; stellar mass $M_*$ from the \emph{2MASS} K-band luminosity; baryon mass $M_{TF}$ from the rotation velocity $v_{rot}$ and the baryon Tully-Fisher relation. The sample covers a broad range of galaxy properties; e.g., about three orders of magnitude in the SFR and more than two orders of magnitude in the stellar mass. Various subsamples are further defined for comparison: starburst (13 galaxies) v.s. non-starburst (29); clustered (17) v.s. field (35); early-type (24) v.s. late-type (29) disc galaxies.

The \emph{Chandra} data of the sample galaxies are calibrated and analyzed in a uniform manner. For example, the diffuse X-ray spectrum is extracted for each galactic corona, typically from a field within a vertical distance from the galactic disc $|z|\lesssim5~h_{exp}$ (see Table~\ref{table:CombinedPara} for the average vertical extension of this field), where $h_{exp}$ is the exponential scale height of the vertical brightness profile. Both the resolved and unresolved stellar contributions are excised/subtracted or quantitatively estimated. The coronal luminosity is then calculated in the 0.5-2~keV band and corrected for removed source areas, CCD edges, and galactic discs. Similar corrections are made also for derived hot gas properties, e.g., density, mass, thermal energy, and radiative cooling timescale. For galaxies with sufficient X-ray counting statistics, spectral constraints are further obtained on the thermal (temperature) and chemical (measured in the form of Fe/O abundance ratio) states of the coronal gas. Galaxies with distinct multi-wavelength characteristics, which may affect the coronal properties, are individually noted.

Furthermore, discussions are given about various potential complications which may cause biases to the characterization of the coronae. In particular, the effects caused by uncertainties in the thermal and chemical states of the hot gas are investigated. We find that a simple 1-T model with fixed abundance ratio plus stellar contributions is typically sufficient to characterize the overall properties of the coronal emission. The inclination angle range of our sample galaxies may cause a small systematical bias due to the potential contamination from the disc emission/absorption in moderately inclined galaxies. But this bias cannot play a significant role in determining the coronal properties. Several galaxies in our sample host AGNs; some of them may even be responsible for some distinct diffuse X-ray features. But in general, the presence of these diffuse X-ray features related to AGN does not significantly affect our measurements of the global coronal properties. The CXE may significantly contribute to the soft X-ray emission in some of the sample galaxies, either active in SF or not. In particular, the unusually high X-ray luminosity of the tidally disturbed gas-rich Virgo cluster member NGC~4438 could well be due to the CXE. Generally, these complications should not prevent a meaningful correlation analysis, which can yield further useful insights into the nature of the X-ray emission and its relationship to galaxy properties.

\subsection{Combination of Different Parameters}\label{PaperIIsubsec:CombinationPara}

While relevant parameters characterizing galaxy properties have been described in Paper~I, we here further define a few useful parameter combinations in our correlation analysis (Table~\ref{table:CombinedPara}).

(1) Mass ratio ($M_{TF}/M_*$). $M_{TF}$ is measured from the rotation velocity and thus traces the central gravity, as well as the total cool baryon, of a galaxy. This mass ratio thus characterizes the nonlinearity of the gravitational to stellar mass relation. We will further discuss its physical meaning in \S\ref{PaperIIsubsec:DiscussionScaling}.

(2) Total SN mechanical energy injection rate ($\dot{E}_{SN}=\dot{E}_{SN(Ia)}+\dot{E}_{SN(CC)}$). We estimate this rate from the two types of SNe: Type~Ia ($\dot{E}_{SN(Ia)}$) from the stellar mass \citep{Mannucci05} and CC ($\dot{E}_{SN(CC)}$) from the SFR \citep{Heckman90}, using the following explosion rates:
\begin{equation}\label{equi:Iarate}
\nu_{Ia}=0.044 (M_*/10^{10}{\rm~M_\odot}) \rm~century^{-1},
\end{equation}
\begin{equation}\label{equi:CCrate}
\nu_{CC}=0.77 ({\rm SFR}/{\rm M_\odot~yr^{-1}}) \rm~century^{-1},
\end{equation}
and assuming $10^{51}\rm~erg$ per SN. The dependence of the Ia SN rate on stellar population is accounted for in the calculation of stellar mass, where we have adopted the color-dependent mass-to-light ratio from \citet{Bell01} (Paper~I).

(3) Surface rate of CC SNe ($F_{SN(CC)}$). According to the standard superbubble theory \citep{MacLow88}, the blowout of galactic disc gas is determined by the energy injection rate per unit disc area. There exists a critical surface rate for superbubble blowout from a galactic disc \citep{Strickland04b}. Here we consider only CC SNe, which contribute to the disc blowout via coherent energy input, in both space and time. The disc area is calculated using the optical diameter of the whole galactic disc $D_{25}$, since the collective energy injection from star clusters distributed widely across the disk is shown to be effective \citep{Strickland04b}.

(4) X-ray radiation efficiency ($\eta\equiv L_X/\dot{E}_{SN}$) defined as the fraction of the SN energy radiated in soft X-ray.

\begin{deluxetable}{lccccccccccccc}
\centering
\tiny 
  \tabletypesize{\tiny}
  \tablecaption{Parameters of the Sample Galaxies}
  \tablewidth{0pt}
  \tablehead{
 \colhead{Name} & \colhead{$M_{TF}/M_*$} & \colhead{$\dot{E}_{SN}$} & \colhead{$\dot{E}_{SN(Ia)}$} & \colhead{$\dot{E}_{SN(CC)}$} & \colhead{$F_{SN(CC)}$} & \colhead{$\eta$} & \colhead{$h_{avg}$} \\
   &  & ($10^{38}\rm erg/s$) & ($10^{38}\rm erg/s$) & ($10^{38}\rm erg/s$) & ($\rm SN/Myr/kpc^2$) & (\%) & (kpc) \\
   & (1) & (2) & (3) & (4) & (5) & (6) & (7) 
}
\startdata
IC2560 & $6.08_{-0.35}^{+0.36}$ & $5.16_{-0.74}^{+0.77}$ & $0.150\pm0.004$ & $5.01\pm0.78$ & $22.1\pm3.4$ & $2.23_{-0.29}^{+0.44}$ & 12.5 \\
M82 & $0.93\pm0.09$ & $19.1\pm1.1$ & $0.278\pm0.004$ & $18.8\pm1.1$ & $591\pm35$ & $0.62_{-0.03}^{+0.04}$ & 4.4 \\
NGC0024 & $3.20_{-0.17}^{+0.18}$ & $0.28_{-0.14}^{+0.02}$ & $0.021\pm0.0005$ & $0.26_{-0.14}^{+0.02}$ & $3.97_{-2.15}^{+0.25}$ & $0.60_{-0.11}^{+5.41}$ & 2.5 \\
NGC0520 & $0.053\pm0.006$ & $29.2\pm3.7$ & $0.516\pm0.009$ & $28.6\pm3.9$ & $104\pm14$ & $0.067_{-0.020}^{+0.016}$ & 10.8 \\
NGC0660 & $0.69\pm0.05$ & $17.8_{-2.2}^{+2.1}$ & $0.416\pm0.006$ & $17.4\pm2.2$ & $182\pm23$ & $0.066_{-0.012}^{+0.020}$ & 4.7 \\
NGC0891 & $1.75\pm0.15$ & $6.68_{-0.82}^{+0.79}$ & $0.688\pm0.010$ & $5.99\pm0.82$ & $16.9\pm2.3$ & $0.57_{-0.06}^{+0.08}$ & 7.1 \\
NGC1023 & $0.14\pm0.02$ & - & $0.942\pm0.014$ & - & - & - & 4.4 \\
NGC1380 & - & $2.83_{-0.49}^{+0.06}$ & $1.72\pm0.03$ & $1.11_{-0.55}^{+0.05}$ & $5.59_{-2.77}^{+0.25}$ & $1.42_{-0.13}^{+0.29}$ & 4.3 \\
NGC1386 & - & $2.46\pm0.19$ & $0.254\pm0.004$ & $2.20\pm0.19$ & $34.6\pm3.0$ & $0.61_{-0.08}^{+0.07}$ & 6.7 \\
NGC1482 & $0.53\pm0.12$ & $16.3_{-1.8}^{+1.9}$ & $0.319\pm0.005$ & $15.9\pm1.9$ & $325\pm38$ & $0.46_{-0.06}^{+0.07}$ & 5.1 \\
NGC1808 & $0.33\pm0.05$ & $19.5\pm1.4$ & $0.527\pm0.008$ & $19.0\pm1.5$ & $202\pm16$ & $0.13\pm0.02$ & 3.2 \\
NGC2787 & $1.51_{-0.36}^{+0.37}$ & $1.09_{-0.50}^{+0.01}$ & $0.465\pm0.005$ & $0.63_{-0.57}^{+0.01}$ & $16.9_{-15.2}^{+0.2}$ & $0.16_{-0.14}^{+0.65}$ & 3.2 \\
NGC2841 & $3.65\pm0.35$ & $2.56_{-0.33}^{+0.12}$ & $1.37\pm0.02$ & $1.19_{-0.34}^{+0.12}$ & $5.94_{-1.73}^{+0.62}$ & $0.77_{-0.12}^{+0.17}$ & 6.4 \\
NGC3079 & $2.82_{-0.24}^{+0.23}$ & $15.2\pm1.3$ & $0.414\pm0.006$ & $14.8\pm1.4$ & $38.4\pm3.7$ & $0.57_{-0.05}^{+0.06}$ & 12.5 \\
NGC3115 & $0.12\pm0.02$ & - & $0.947\pm0.014$ & - & - & - & 1.8 \\
NGC3198 & $2.35\pm0.23$ & $1.66_{-0.30}^{+0.13}$ & $0.146\pm0.003$ & $1.52_{-0.30}^{+0.13}$ & $8.21_{-1.65}^{+0.73}$ & $1.01_{-0.12}^{+0.26}$ & 8.3 \\
NGC3384 & $2.8_{-0.27}^{+0.28}\times10^{-4}$ & - & $0.574\pm0.008$ & - & - & - & 7.3 \\
NGC3412 & - & - & $0.232\pm0.003$ & - & - & - & 2.8 \\
NGC3521 & $1.71_{-0.17}^{+0.16}$ & $6.18_{-0.96}^{+0.57}$ & $0.984\pm0.014$ & $5.20_{-0.98}^{+0.59}$ & $28.4_{-5.4}^{+3.2}$ & $0.43_{-0.04}^{+0.09}$ & 10.3 \\
NGC3556 & $1.71_{-0.12}^{+0.11}$ & $4.05_{-0.34}^{+0.36}$ & $0.224\pm0.004$ & $3.82\pm0.37$ & $100.0\pm9.7$ & $0.32\pm0.07$ & 4.9 \\
NGC3628 & $1.35\pm0.08$ & $12.7_{-1.2}^{+1.3}$ & $0.939\pm0.014$ & $11.8\pm1.3$ & $26.7\pm3.0$ & $0.30\pm0.04$ & 17.5 \\
NGC3877 & $1.65\pm0.15$ & $1.99\pm0.19$ & $0.243\pm0.004$ & $1.74\pm0.20$ & $14.5\pm1.6$ & $0.22\pm0.11$ & 3.6 \\
NGC3955 & $0.29\pm0.03$ & $5.25_{-0.63}^{+0.62}$ & $0.180\pm0.004$ & $5.07\pm0.66$ & $32.9\pm4.3$ & $0.21_{-0.11}^{+0.08}$ & 5.2 \\
NGC3957 & - & $1.80_{-1.08}^{+0.08}$ & $0.155\pm0.002$ & $1.64_{-1.14}^{+0.08}$ & $9.78_{-6.78}^{+0.47}$ & $1.10_{-6.67}^{+2.85}$ & 4.4 \\
NGC4013 & $1.12\pm0.11$ & $3.98_{-0.71}^{+0.42}$ & $0.627\pm0.009$ & $3.36_{-0.73}^{+0.44}$ & $18.7_{-4.1}^{+2.5}$ & $0.58_{-0.07}^{+0.14}$ & 6.0 \\
NGC4111 & $0.063\pm0.013$ & - & $0.421\pm0.006$ & - & - & - & 3.8 \\
NGC4217 & $1.32_{-0.10}^{+0.11}$ & - & $0.592\pm0.009$ & - & - & - & 4.5 \\
NGC4244 & $4.67\pm0.38$ & $0.060_{-0.033}^{+0.001}$ & $0.012\pm0.0002$ & $0.047_{-0.035}^{+0.001}$ & $0.45_{-0.33}^{+0.01}$ & $1.81_{-5.39}^{+5.09}$ & 1.9 \\
NGC4251 & - & - & $0.590\pm0.006$ & - & - & - & 3.0 \\
NGC4342 & - & - & $0.176\pm0.002$ & - & - & - & 12.7 \\
NGC4388 & $2.62_{-0.27}^{+0.26}$ & $5.67\pm0.53$ & $0.227\pm0.004$ & $5.44\pm0.54$ & $30.5\pm3.0$ & $1.29_{-0.18}^{+0.25}$ & 10.3 \\
NGC4438 & $1.19_{-0.24}^{+0.23}$ & $1.73_{-0.34}^{+0.11}$ & $0.441\pm0.006$ & $1.29_{-0.36}^{+0.13}$ & $3.52_{-0.98}^{+0.35}$ & $4.78_{-0.44}^{+1.29}$ & 11.8 \\
NGC4501 & $2.69\pm0.26$ & $7.40_{-0.82}^{+0.80}$ & $1.13\pm0.02$ & $6.27\pm0.86$ & $16.1\pm2.2$ & $1.87_{-0.35}^{+0.44}$ & 6.7 \\
NGC4526 & $0.22_{-0.05}^{+0.04}$ & $4.38_{-0.65}^{+0.34}$ & $1.66\pm0.02$ & $2.72_{-0.67}^{+0.36}$ & $9.04_{-2.24}^{+1.19}$ & $0.43_{-0.09}^{+0.13}$ & 4.5 \\
NGC4565 & $2.71_{-0.22}^{+0.21}$ & $2.37_{-0.59}^{+0.26}$ & $0.735\pm0.010$ & $1.63_{-0.62}^{+0.27}$ & $2.24_{-0.86}^{+0.37}$ & $0.46_{-0.06}^{+0.16}$ & 9.1 \\
NGC4569 & $2.20\pm0.37$ & $1.48_{-0.24}^{+0.14}$ & $0.290\pm0.005$ & $1.19_{-0.25}^{+0.15}$ & $6.98_{-1.49}^{+0.86}$ & $1.58_{-0.55}^{+0.44}$ & 6.9 \\
NGC4594 & $3.52_{-0.35}^{+0.33}$ & $2.84_{-0.23}^{+0.08}$ & $2.16\pm0.03$ & $0.68_{-0.25}^{+0.07}$ & $4.75_{-1.72}^{+0.49}$ & $1.38_{-0.09}^{+0.13}$ & 13.8 \\
NGC4631 & $1.92_{-0.19}^{+0.17}$ & $4.21_{-0.46}^{+0.48}$ & $0.142\pm0.002$ & $4.07\pm0.49$ & $15.9\pm1.9$ & $0.87_{-0.08}^{+0.11}$ & 7.3 \\
NGC4666 & $1.52\pm0.07$ & $10.4_{-1.0}^{+1.1}$ & $0.568\pm0.008$ & $9.85\pm1.11$ & $76.9\pm8.6$ & $0.83_{-0.33}^{+0.17}$ & 5.6 \\
NGC4710 & $0.96\pm0.09$ & $2.87_{-0.71}^{+0.26}$ & $0.468\pm0.007$ & $2.40_{-0.73}^{+0.27}$ & $21.0_{-6.4}^{+2.4}$ & $0.21_{-0.12}^{+0.09}$ & 2.1 \\
NGC5102 & $2.79_{-0.26}^{+0.27}$ & $0.044_{-0.011}^{+0.001}$ & $0.021\pm0.0004$ & $0.023_{-0.012}^{+0.001}$ & $1.14_{-0.62}^{+0.06}$ & $1.36_{-0.25}^{+0.61}$ & 2.0 \\
NGC5170 & $3.01\pm0.29$ & $2.04_{-0.65}^{+0.09}$ & $0.659\pm0.012$ & $1.38_{-0.70}^{+0.10}$ & $2.04_{-1.03}^{+0.14}$ & $1.61_{-0.40}^{+1.08}$ & 9.7 \\
NGC5253 & $0.43_{-0.07}^{+0.06}$ & $0.83_{-0.06}^{+0.07}$ & $6.8\pm0.21\times10^{-3}$ & $0.82\pm0.07$ & $93.9\pm7.6$ & $0.22\pm0.02$ & 1.3 \\
NGC5422 & - & - & $0.646\pm0.011$ & - & - & - & 2.2 \\
NGC5746 & $2.52_{-0.26}^{+0.25}$ & $4.33_{-0.77}^{+0.16}$ & $1.99\pm0.03$ & $2.34_{-0.85}^{+0.16}$ & $3.46_{-1.26}^{+0.24}$ & $0.40_{-0.22}^{+0.18}$ & 6.1 \\
NGC5775 & $0.89\pm0.11$ & $17.8_{-1.6}^{+1.7}$ & $0.916\pm0.012$ & $16.9\pm1.7$ & $82.3\pm8.1$ & $0.57_{-0.07}^{+0.08}$ & 8.2 \\
NGC5866 & - & $2.77\pm0.16$ & $0.771\pm0.011$ & $2.00\pm0.17$ & $10.2\pm0.9$ & $0.49_{-0.09}^{+0.08}$ & 5.5 \\
NGC6503 & $0.48\pm0.05$ & $0.40\pm0.03$ & $0.044\pm0.001$ & $0.36\pm0.03$ & $17.5\pm1.7$ & $0.38_{-0.05}^{+0.06}$ & 2.1 \\
NGC6764 & $2.01\pm0.19$ & $6.81_{-0.58}^{+0.59}$ & $0.139\pm0.005$ & $6.67\pm0.62$ & $74.1\pm6.9$ & $2.75_{-1.10}^{+0.83}$ & 1.5 \\
NGC7090 & $4.47_{-0.41}^{+0.44}$ & $0.37_{-0.06}^{+0.03}$ & $0.021\pm0.0004$ & $0.35_{-0.06}^{+0.03}$ & $6.26_{-1.16}^{+0.58}$ & $0.12\pm0.10$ & 1.9 \\
NGC7457 & - & - & $0.170\pm0.004$ & - & - & - & 7.1 \\
NGC7582 & $0.95\pm0.05$ & $32.3\pm3.8$ & $0.943\pm0.015$ & $31.3\pm4.1$ & $58.2\pm7.7$ & $0.32_{-0.06}^{+0.07}$ & 3.2 \\
NGC7814 & $1.65_{-0.20}^{+0.19}$ & - & $0.979\pm0.014$ & - & - & - & 3.8
\enddata
\tablecomments{\scriptsize Galaxy parameters from the combination of the existing parameters in Paper~I (see \S\ref{PaperIIsubsec:CombinationPara}): (1) dynamical-to-photometric mass ratio; (2-4) total, Type~Ia, and core collapsed SN energy injection rate; (5) surface core collapsed SN rate; (6) X-ray radiation efficiency; (7) average vertical extensions (of the positive and negative side, see Paper~I for details) of the regions used to calculate the coronal luminosity. ``-'' mean no relevant data for calculating the entry.
}\label{table:CombinedPara}
\end{deluxetable}

\section{Correlation Analysis}\label{PaperIIsec:correlation}

\subsection{Method}\label{PaperIIsubsec:Method}

In the following correlation analysis, we utilize the Spearman's rank order coefficient ($r_s$; by definition, $-1<r_s<1$) to describe the goodness of a correlation. This simple goodness description works for any pair of variables that show a monotonic correlation, insensitive to its exact form. We consider $|r_s|\gtrsim0.6$ or $0.3\lesssim|r_s|\lesssim0.6$ as a tight or weak correlation, and $|r_s|\lesssim0.3$ as no correlation. We further characterize important tight correlations with simple relations, together with the corresponding dispersion measurements [the root mean square ($rms$) around the fitted relations]. To fit a tight correlation with expressions and to estimate the uncertainties of $r_s$ and $rms$, we first generate 1000 bootstrap-with-replacement samples of the primary data points (those actually measured, not parameter combinations such as a ratio) and then resample each data point, assuming a normal distribution with the expected value and errors (on all sides) as measured from the original data in Paper I. For each re-generated sample, we fit the data with the same expression to get its parameters and measure both $r_s$ and $rms$ again. These measurements are then rank-ordered; their 68\% percentiles around the original fitting parameters, $r_s$, and $rms$ values are finally used as the estimates of their 1$\sigma$ uncertainties, which account for the systematic dispersion among the original data points as well as the uncertainties in their individual measurements. The correlations invoking parameter combinations are also calculated using the same re-generated data (e.g., parameter ratios are re-calculated from each data set). All errors quoted in this paper are at the $1~\sigma$ confidence level.

\subsection{Correlations of Coronal X-ray Luminosity with Other Galaxy Parameters}\label{PaperIIsubsec:BasicScaling}

\begin{figure}[!h]
\begin{center}
\epsfig{figure=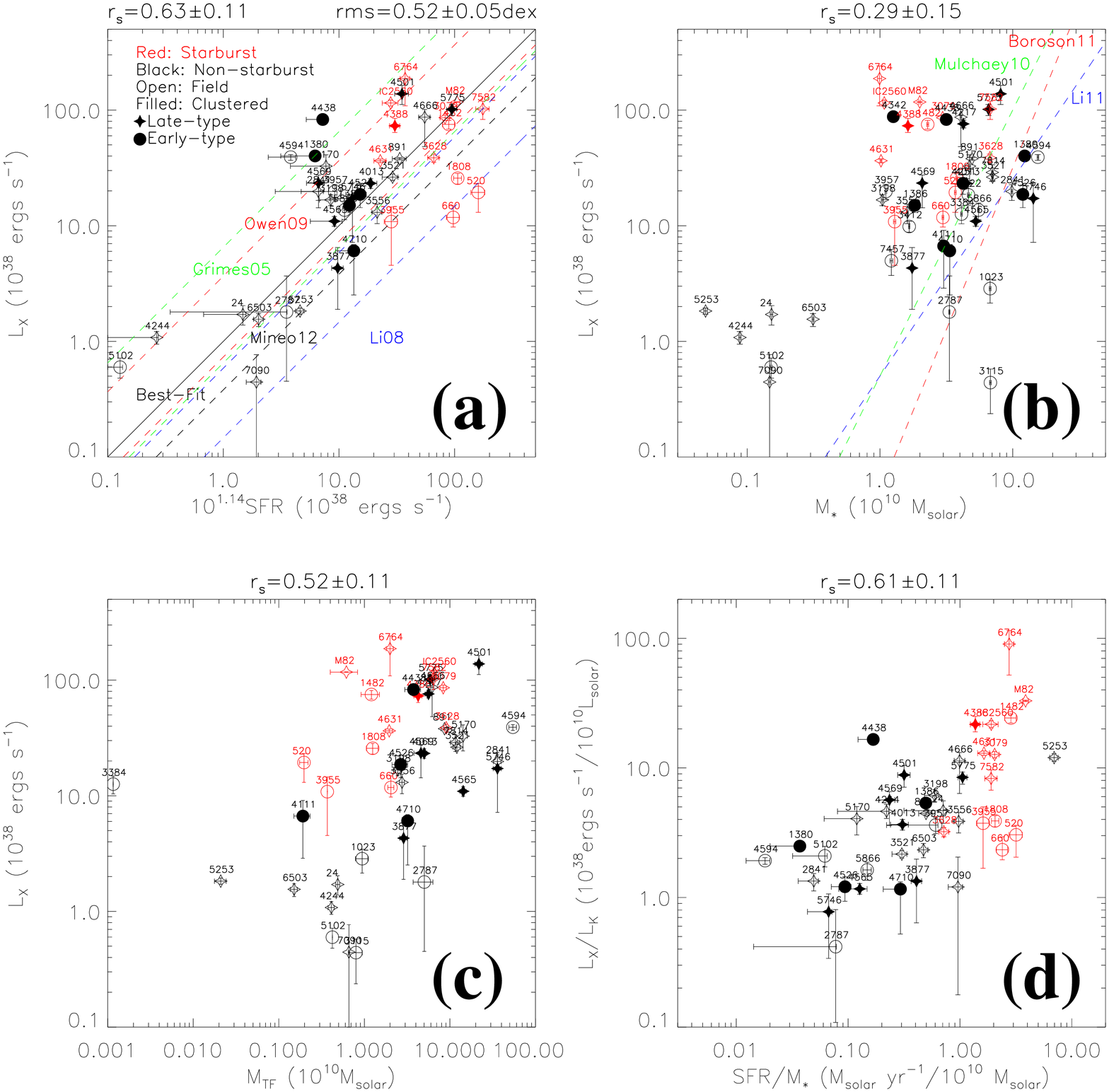,width=1.0\textwidth,angle=0, clip=}
\caption{0.5-2~keV luminosity of the galactic corona ($L_X$) vs. various galaxy properties (see Paper~I for details). The solid line in (a) shows the best-fit linear function (Eq.~\ref{equi:SFRLX}), while the dashed lines mark the ranges from \citet{Grimes05} (green), \citet{Li08} [blue; including \citet{Strickland04a,Strickland04b}'s sample], \citet{Owen09} (red), and \citet{Mineo12} (black; only the best-fit value of the thermal component), respectively. The colored dashed lines in (b) show the relations of S0 galaxies \citep{Li11} and elliptical galaxies \citep{Mulchaey10,Boroson11}. The symbols used in these panels are noted in (a). Also given are the Spearman's rank order coefficients ($r_s$) as well as the $rms$ scatter around the best-fit relation.}\label{fig:correlationLX}
\end{center}
\end{figure}

We first examine the correlations of $L_X$ with various directly measured galaxy properties (e.g., SFR, $M_*$, and $M_{TF}$) and characterize the most significant ones with simple scaling relations. Fig.~\ref{fig:correlationLX}a shows a tight correlation between $L_X$ and SFR ($r_s=0.63\pm0.11$), which we characterize linearly as:
\begin{equation}\label{equi:SFRLX}
L_X({10^{38}\rm ergs~s^{-1}})=10^{(1.14_{-0.37}^{+0.26})}{\rm SFR}({\rm M_\odot~yr^{-1}}).
\end{equation}
The normalization of the relation is somewhat different from those (Fig.~\ref{fig:correlationLX}a) obtained in previous X-ray studies of face-on spiral galaxies (e.g., \citealt{Owen09}) or high latitude halos of starburst galaxies (e.g., \citealt{Strickland04b}). These differences are \emph{not} particularly significant, however, because of the large scatter of the data points around the relation ($rms=0.52\pm0.05\rm~dex$), as well as the different point source removal, absorption correction, photometry area, and inclination angle ranges of these various galaxy samples \citep{Mineo12}. Interestingly, the $L_X-{\rm SFR}$ correlation is not significant for the starburst ($r_s=-0.03\pm0.30$), early-type ($r_s=0.16\pm0.30$), and clustered ($r_s=0.38\pm0.29$) subsamples. However, a tight correlation ($r_s=0.92\pm0.10$) is still found for the eight galaxies included in the study by \citet{Strickland04a,Strickland04b}, which include five starburst ones (M82, NGC~1482, NGC~3079, NGC~3628, NGC~4631) and three non-starburst others (NGC~891, NGC~4244, NGC~6503) (NGC~253 and NGC~4945 are not included in our sample because of their large angular sizes).

As discussed in Paper~I, various inclination angles of the present sample galaxies may affect the measurements of hot gas parameters. We check the correlations for highly ($\geq80^\circ$) and moderately ($<80^\circ$) inclined galaxies separately. No significant differences (in terms of the goodness of the correlation or the fitted relations) between these two subsamples are found. For example, for the $L_X-{\rm SFR}$ relation, the highly and moderately inclined subsamples have $r_s=0.64\pm0.18$ and $0.55\pm0.18$, while the fitted normalizations of the linear relations are $10^{(1.16_{-0.23}^{+0.16})}$ and $10^{(1.10_{-0.39}^{+0.43})}$, respectively. It thus seems that the required inclination corrections or their uncertainties are generally small, compared to the intrinsic diversity of the coronal properties. Therefore, we do not separate the highly and moderately inclined subsamples in the subsequent analysis.

The correlations of $L_X$ with other galaxy parameters are generally much weaker if present. $L_X$ and $M_*$ are at most weakly correlated for the whole sample ($r_s=0.29\pm0.15$; Fig.~\ref{fig:correlationLX}b), although excluding starburst galaxies makes the correlation tighter ($r_s=0.50\pm0.14$). In the $L_X$~vs.~$M_*$ plot (Fig.~\ref{fig:correlationLX}b), the distribution of the disc galaxies is significantly flatter than that of massive elliptical galaxies (e.g., \citealt{Mulchaey10,Boroson11}) or S0 galaxies \citep{Li11}. Disc galaxies also tend to be more X-ray luminous than earlier-type ones of similar stellar masses (see also \S\ref{PaperIIsubsec:DiscussionElliptical} for more discussions). Similarly, the correlation between $L_X$ and $M_{TF}$ is weak for the whole sample ($r_s=0.52\pm0.11$), but is tighter for non-starburst galaxies only ($r_s=0.68\pm0.08$) (Fig.~\ref{fig:correlationLX}c).

Generally, the correlations of $L_X$ with SFR, $M_*$, and $M_{TF}$ are good for \emph{non-starburst} galaxies ($r_s=0.61\pm0.15$, $0.50\pm0.14$, and $0.68\pm0.08$, respectively). However, although the relative position of starburst and non-starburst galaxies are similar in the $L_X-M_*$ and $L_X-M_{TF}$ relations (starburst galaxies tend to be offset to higher $L_X$), it is significantly different in the $L_X-{\rm SFR}$ relation, where the two subsamples can be described with an identical linear relation.

We also compare the specific SFR (per stellar mass) to the specific X-ray luminosity (per K-band luminosity) (Fig.~\ref{fig:correlationLX}d). The correlation between these two parameters are distance-independent and are less affected (than the $L_X-{\rm SFR}$ relation) by the intrinsic scaling of the SFR with the galaxy masses (\S\ref{PaperIIsec:Introduction}). The overall correlation is good ($r_s=0.61\pm0.11$) and even holds for low-mass galaxies with weak coronal X-ray emission.

\begin{figure}[!h]
\begin{center}
\epsfig{figure=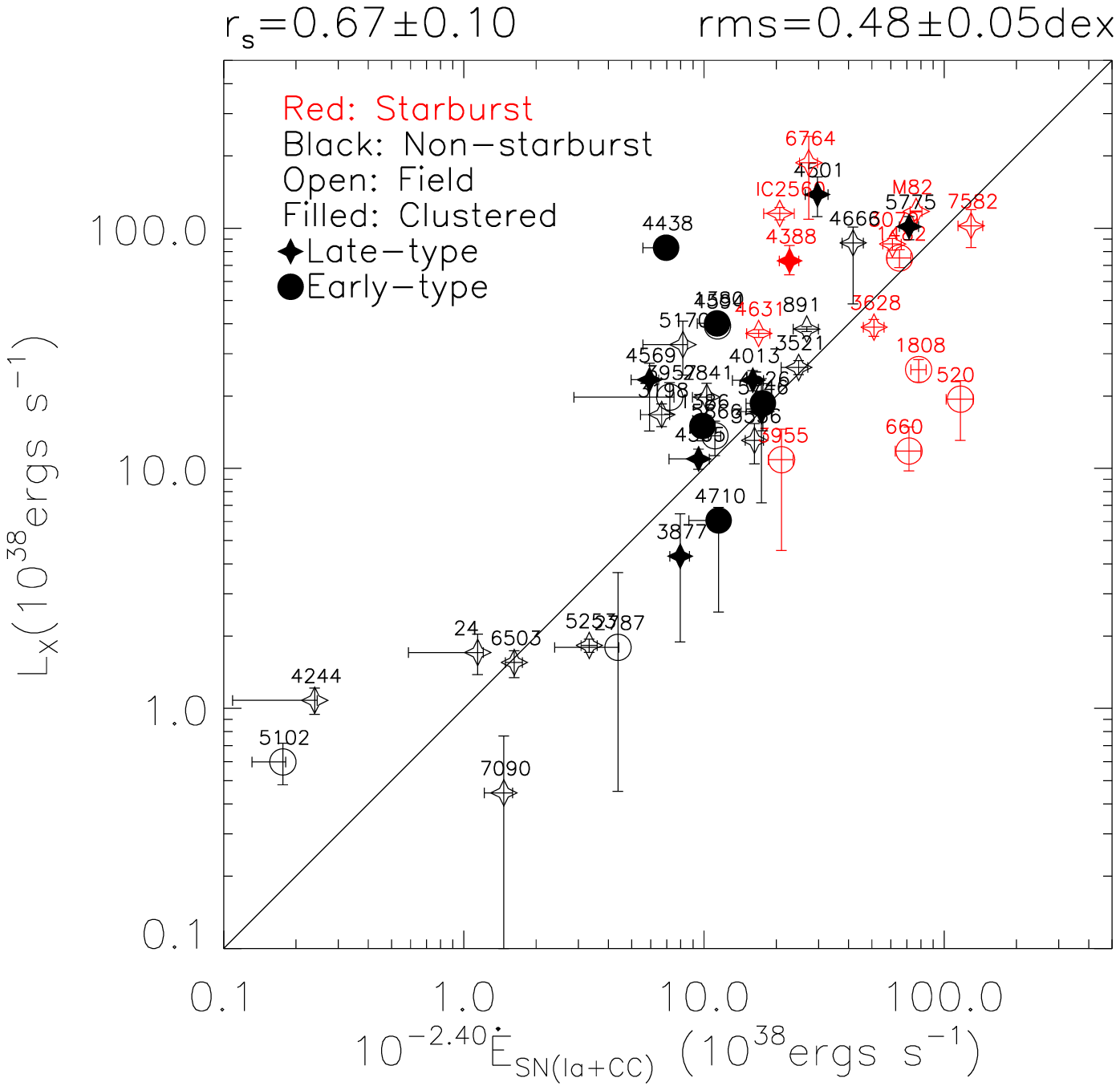,width=0.55\textwidth,angle=0, clip=}
\caption{$L_X$ plotted against the expected total SN energy input rate $\dot{E}_{SN(CC+Ia)}$.}\label{fig:ESN}
\end{center}
\end{figure}

We next replace the SFR in the above correlations with the expected total SN mechanical energy input rate ($\dot{E}_{SN}$). In normal galaxies with little SF, Type~Ia SNe are expected to play an important role in gas heating (e.g., \citealt{LiZ07a}). Fig.~\ref{fig:ESN} shows an improved correlation, which can be characterized as:
\begin{equation}\label{equi:SNELX}
L_{X}=10^{-(2.40_{-0.09}^{+0.10})}\dot{E}_{SN(Ia+CC)}.
\end{equation}
The improvement is the most significant for non-starburst galaxies ($r_s$ changes from $0.61\pm0.15$ to $0.70\pm0.12$; $rms$ is reduced from $0.47\pm0.06$ to $0.40\pm0.07$). For some individual galaxies (e.g., NGC~4594; \citealt{LiZ07a}), Ia SNe can contribute $\gtrsim75\%$ of the total energy input rate (Table~\ref{table:CombinedPara}).

We also test the possibility that Ia and CC SNe may have different X-ray radiation efficiencies. The two efficiencies are poorly constrained in a joint fit including the two components (of Ia and CC SNe) and are consistent with each other within errors. Therefore, we will not separate the two types of SNe in the following discussions.

\subsection{Multi-Parameter Joint Correlations}\label{PaperIIsubsec:JointScaling}

\begin{figure}[!h]
\begin{center}
\epsfig{figure=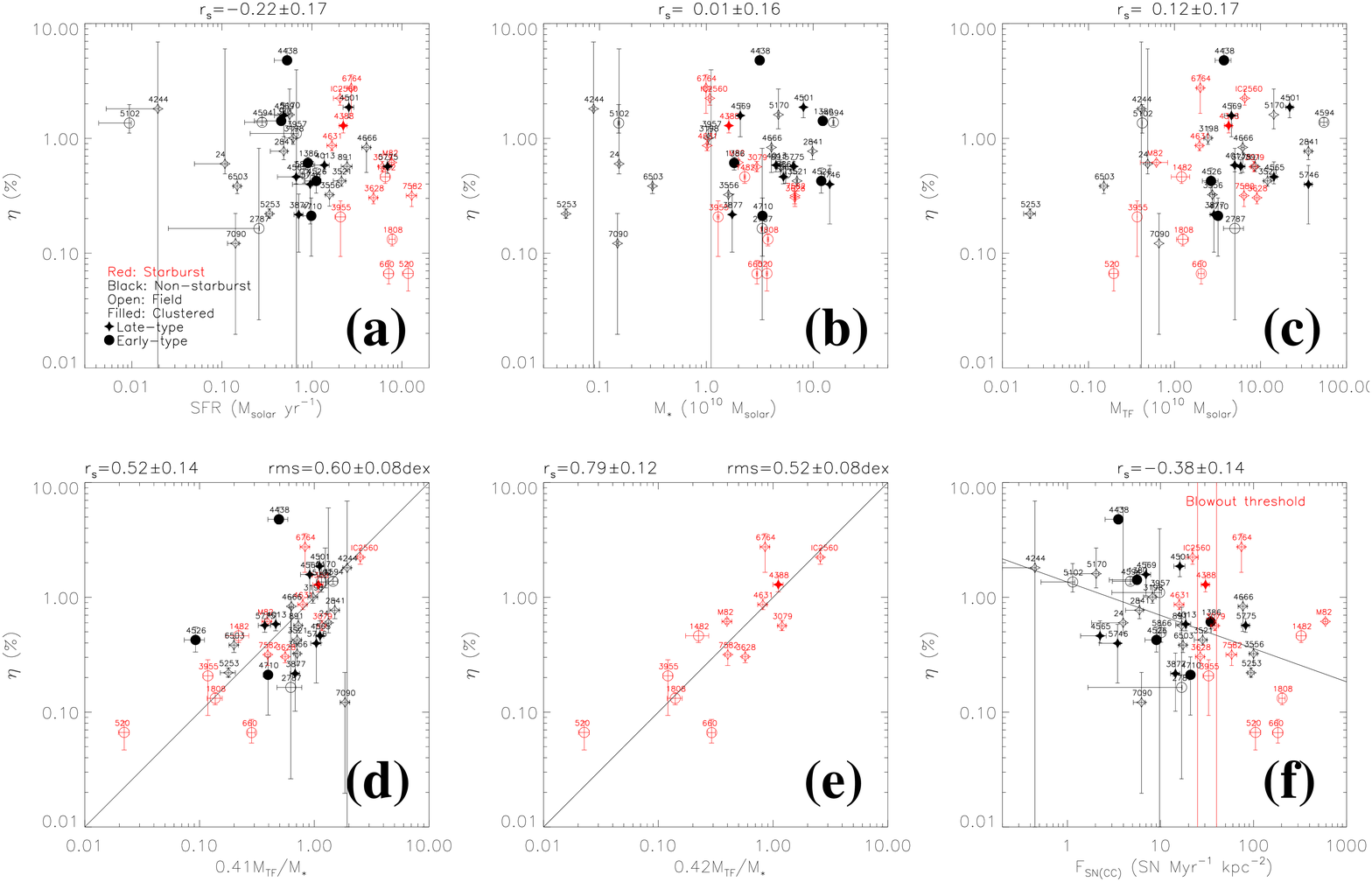,width=1.0\textwidth,angle=0, clip=}
\caption{X-ray radiation efficiency ($\eta$) plotted against various galaxy properties. (e) is the same as (d), but includes only starburst galaxies for clarify. The black lines in (d-f) show the fitted relations as described in the text. The two red lines in (f) mark the blowout threshold obtained by \citet{Strickland04b}; $F_{SN}=25\rm~SN~Myr^{-1}~kpc^{-2}$ is a model prediction assuming an ambient medium density $n_0\sim1\rm~cm^{-3}$, thermal pressure $P/k\sim10^4\rm~K~cm^{-3}$, and the feedback efficiency $\epsilon\Upsilon\sim0.14$; $F_{SN}\sim40\rm~SN~Myr^{-1}~kpc^{-2}$ is constrained from observations (see \citealt{Strickland04b} or \S\ref{PaperIIsubsec:DiscussionScaling} for the definition).}\label{fig:efficiency}
\end{center}
\end{figure}

In addition to the primary correlation between $L_X$ and $\dot{E}_{SN}$ (Fig.~\ref{fig:ESN}), we now study the secondary effects, i.e., after taking out the primary $L_X-\dot{E}_{SN}$ correlation, by comparing the X-ray radiation efficiency $\eta$ (\S\ref{PaperIIsubsec:CombinationPara}) to other galaxy properties. As shown in Eq.~\ref{equi:SNELX}, $\eta$ has a mean value of $\sim0.4\%$, suggesting that the bulk of the SN energy is not radiated from the coronae explored here. The large dispersion of $\eta$ ($rms=0.50\pm0.06~\rm dex$) further indicates its potential dependence on other galaxy properties. As expected, Figs.~\ref{fig:efficiency}a,b show no significant global correlation of $\eta$ with SFR and $M_*$, because the definition of $\eta$ has already accounted for the major dependence of $\dot{E}_{SN}$ on these two parameters. Similar to the $\eta-{\rm SFR}$ or $\eta-M_*$ relations, there is no correlation between $\eta$ and $M_{TF}$ (Fig.~\ref{fig:efficiency}c). However, a weak positive correlation of $\eta$ with $M_{TF}/M_*$ ($r_s=0.52\pm0.14$) is apparent (Fig.~\ref{fig:efficiency}d), which can be characterized as:
\begin{equation}\label{equi:EtaMratio}
\eta=(0.41_{-0.12}^{+0.13}\%) M_{TF}/M_*.
\end{equation}
Although the definition of both $\eta$ and $M_{TF}/M_*$ include $M_*$, this seems insufficient to explain the correlation in Fig.~\ref{fig:efficiency}d. In starburst galaxies, Type~Ia SNe (whose rate is proportional to $M_*$) make a negligible contribution to the total energy input rate (and so $\eta$), but $\eta$ of these galaxies shows a much better correlation ($r_s=0.79\pm0.12$) than non-starburst ones ($r_s=0.32\pm0.20$) (Fig.~\ref{fig:efficiency}e). Therefore, the $\eta-M_{TF}/M_*$ correlation is partly responsible for the large dispersion in $\eta$ and the $L_X-\dot{E}_{SN}$ correlation.

\begin{figure}[!h]
\begin{center}
\epsfig{figure=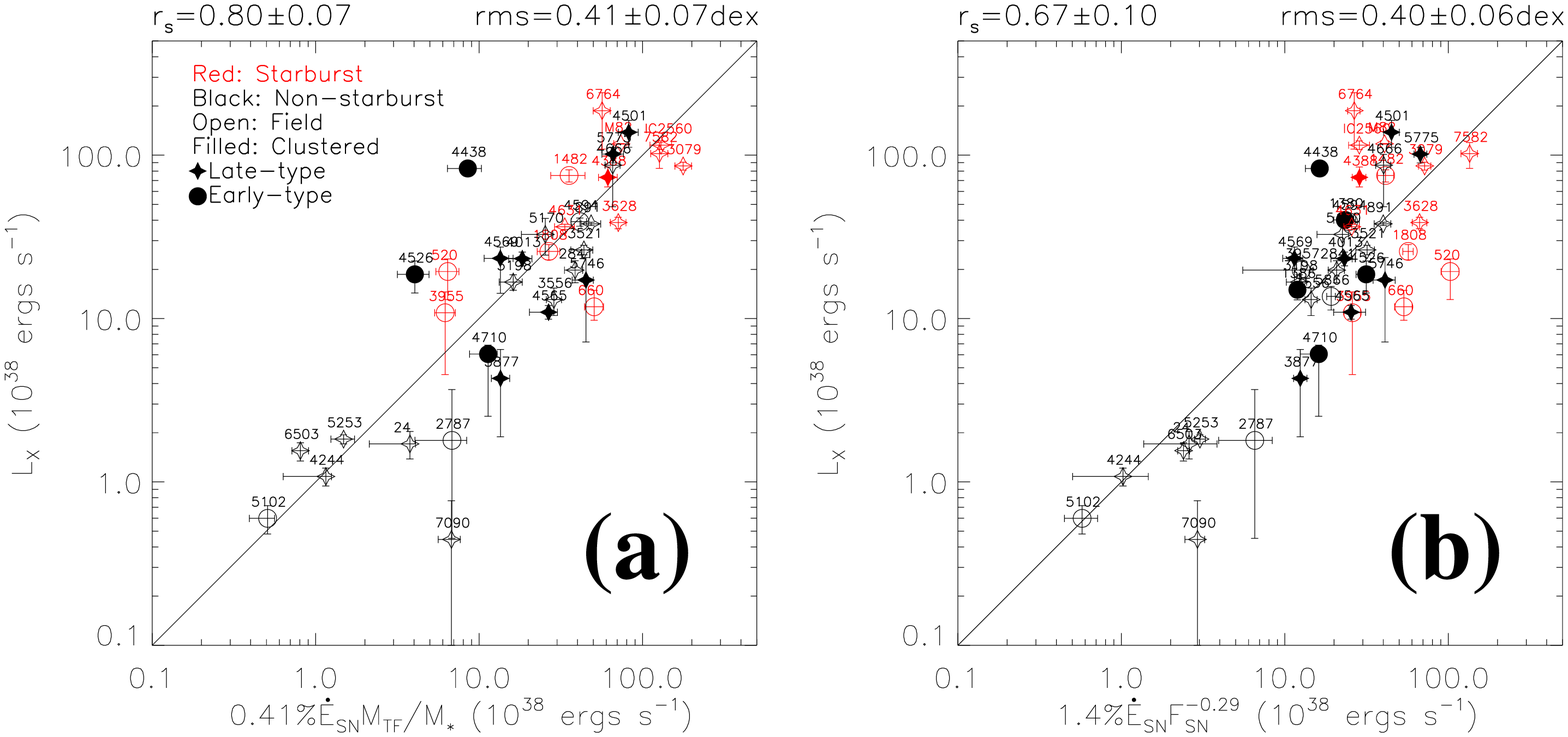,width=1.0\textwidth,angle=0, clip=}
\caption{Joint scaling relations of $L_X$, invoking the best-fit $\eta-M_{TF}/M_*$ and $\eta-F_{SN}$ relations as plotted in Fig.~\ref{fig:efficiency}d,f.}\label{fig:2Dfit}
\end{center}
\end{figure}

For the convenience of comparing the goodness and scatter among different correlations, we rewrite Eq.~\ref{equi:EtaMratio} as a joint scaling relation of (Fig.~\ref{fig:2Dfit}a)
\begin{equation}\label{equi:ESNMTFMstellarLX}
L_X=(0.41_{-0.12}^{+0.13}\%)\dot{E}_{SN(Ia+CC)}M_{TF}/M_*.
\end{equation}
With the inclusion of $M_{TF}/M_*$, the correlations are improved for most of the subsamples. The improvement is the most significant for starburst galaxies ($r_s=0.68\pm0.16$, compared to $-0.03\pm0.30$ for $L_X-{\rm SFR}$ or $-0.05\pm0.30$ for $L_X-\dot{E}_{SN}$).

We also find a weak anti-correlation between $\eta$ and $F_{SN(CC)}$ ($r_s=-0.38\pm0.14$; Fig.~\ref{fig:efficiency}f), which can be characterized as:
\begin{equation}\label{equi:EtaFSNCC}
\eta=(1.4\pm0.5\%)F_{SN(CC)}^{-(0.29\pm0.11)}.
\end{equation}
$F_{SN(CC)}$ is often considered to be directly related to hot gas blowout from galactic discs; two proposed thresholds are shown in Fig.~\ref{fig:efficiency}f: one from empirical assessment and the other from theoretical consideration \citep{Strickland04b}. Both $\eta$ and $F_{SN(CC)}$ depend on the SFR; so the $\eta-F_{SN(CC)}$ anti-correlation may be considered as a secondary non-linear dependence of $L_X$ on SFR (Fig.~\ref{fig:correlationLX}a shows the major linear dependence). In particular, the correlation is much weaker for late-type galaxies ($r_s=-0.16\pm0.20$), in which energy source is mainly contributed by CC SNe, than for early-type ones ($r_s=-0.62\pm0.19$). In a recent absorption line study of 105 galaxies at $0.3<z<1.4$, \citet{Rubin13} find that there is no evidence for a threshold SFR surface density below which galactic winds are not driven. However, they do show a general trend for galaxies with higher SFR surface densities to have a higher detection rate of wind, indicating the importance of $F_{SN(CC)}$ in the generation of galactic winds.

Similarly, Eq.~\ref{equi:EtaFSNCC} can be rewritten as (Fig.~\ref{fig:2Dfit}b):
\begin{equation}\label{equi:ESNFSNCCLX}
L_X=(1.4\pm0.5\%)\dot{E}_{SN(Ia+CC)}F_{SN(CC)}^{-(0.29\pm0.11)}.
\end{equation}
The improvement of the correlation is the most significant for early-type galaxies: e.g., $r_s=0.50\pm0.22$, compared to $0.16\pm0.30$ for $L_X-{\rm SFR}$ and $0.24\pm0.29$ for $L_X-\dot{E}_{SN}$.

\subsection{Other Correlations}\label{PaperIIsubsec:OtherPara}

We compare the characteristic temperatures ($T_X$) from our measurements to those from other recent works: \citet{OSullivan03}'s from the \emph{ROSAT} observations of luminous early-type galaxies; \citet{Grimes05}'s from the \emph{Chandra} observations of nine ultraluminous IR galaxies, six edge-on starburst galaxies, and seven dwarf starburst galaxies; and \citet{Boroson11}'s from the \emph{Chandra} observations of typically X-ray-faint early-type galaxies.

\begin{figure}[!h]
\begin{center}
\epsfig{figure=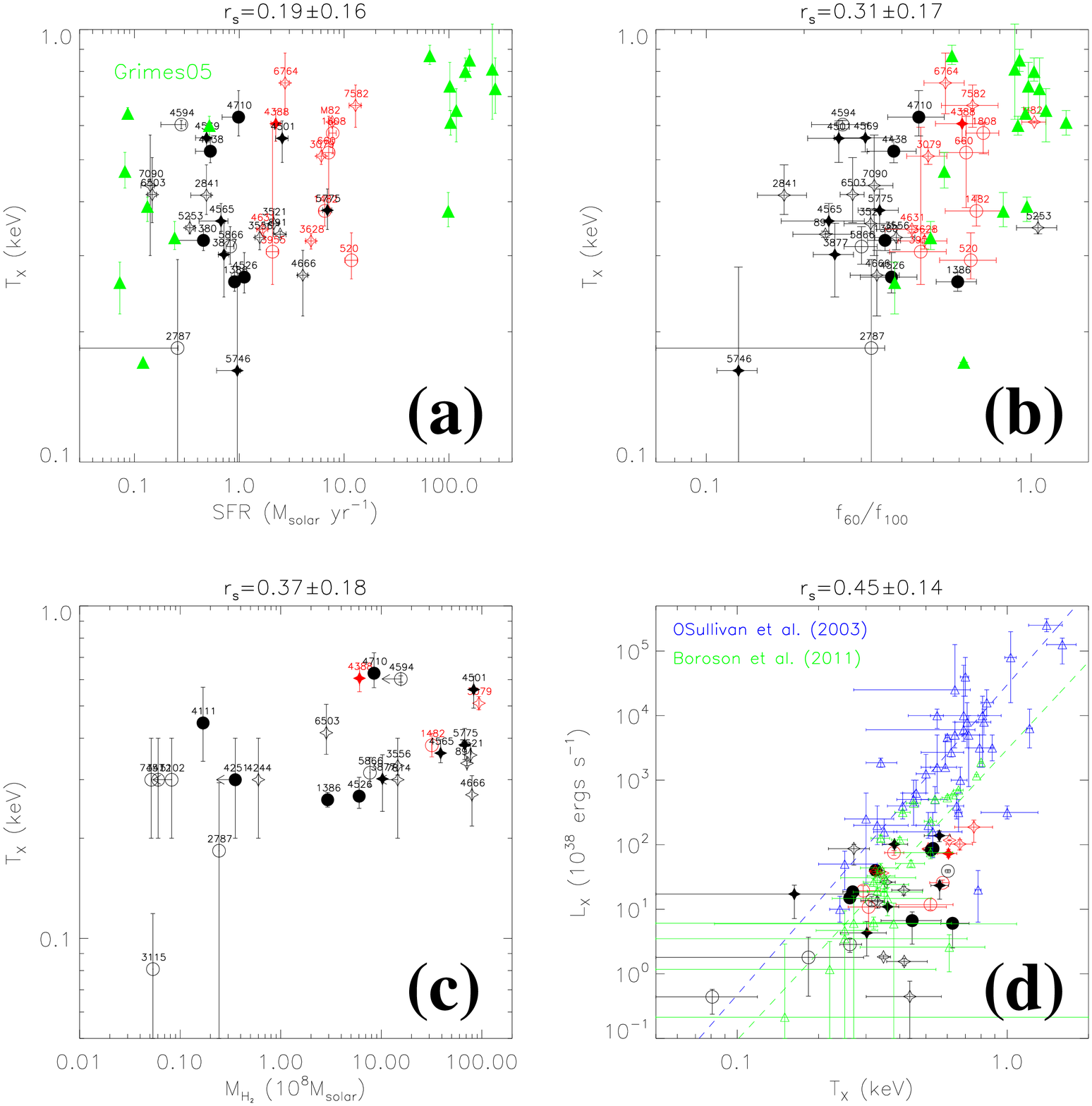,width=1.0\textwidth,angle=0, clip=}
\caption{The characteristic temperature of the 1-T model (Table~6 of Paper~I) vs. various galactic properties: (a) the SFR derived from the IR luminosity (SFR); (b) the far-IR warmth ($f_{60}/f_{100}$); (c) the molecular gas mass ($M_{H_2}$); (d) the soft X-ray luminosity ($L_X$). Symbols are the same as those in Fig.~\ref{fig:correlationLX}. Starburst galaxies from \citet{Grimes05} are included for comparison (green triangles) in (a) and (b). In (c), the blue and green triangles are the data from X-ray bright \citep{OSullivan03} and faint \citep{Boroson11} early-type galaxies, while the dashed lines mark the scaling relations defined with them.}\label{fig:TX}
\end{center}
\end{figure}

We first check the dependence of $T_X$ on the total SFR and the IR warmth, $f_{60}/f_{100}$, which is a proxy of the dust temperature \citep{Strickland04a,Draine07}. Figs.~\ref{fig:TX}a,b show that the correlations are weak and that the data points exhibit large dispersions for all the subsamples. Nevertheless, the most active starburst galaxies tend to have the highest characteristic temperatures \citep{Grimes05}, indicating that the intense SF in the disk could increase the specific energy of the hot gas (\S\ref{PaperIIsubsec:JointScaling}). Cool-hot gas interaction, as suggested in previous studies of both starburst and non-starburst disc galaxies (e.g., \citealt{Strickland00b,Strickland02,Li09,Li11}), can significantly affect the X-ray luminosity, morphology, or the inferred temperature from a corona. However, we find no significant dependence of $T_X$ on the cold gas content [total or specific (per stellar mass) amount of molecular and atomic gas] (e.g., Fig.~\ref{fig:TX}c). One may also expect a tight $L_X-T_X$ correlation for gravitationally heated coronae, as for elliptical galaxies (e.g., \citealt{OSullivan03,Boroson11}). But such a correlation is not seen for most of our disc galaxy subsamples (Fig.~\ref{fig:TX}d), except for starburst galaxies ($r_s=0.64\pm0.23$). Their coronae are unlikely produced by gravitational heating because there is no correlation between $L_X$ and  $M_{TF}$ for these galaxies (\S\ref{PaperIIsubsec:BasicScaling}).

\begin{figure}[!h]
\begin{center}
\epsfig{figure=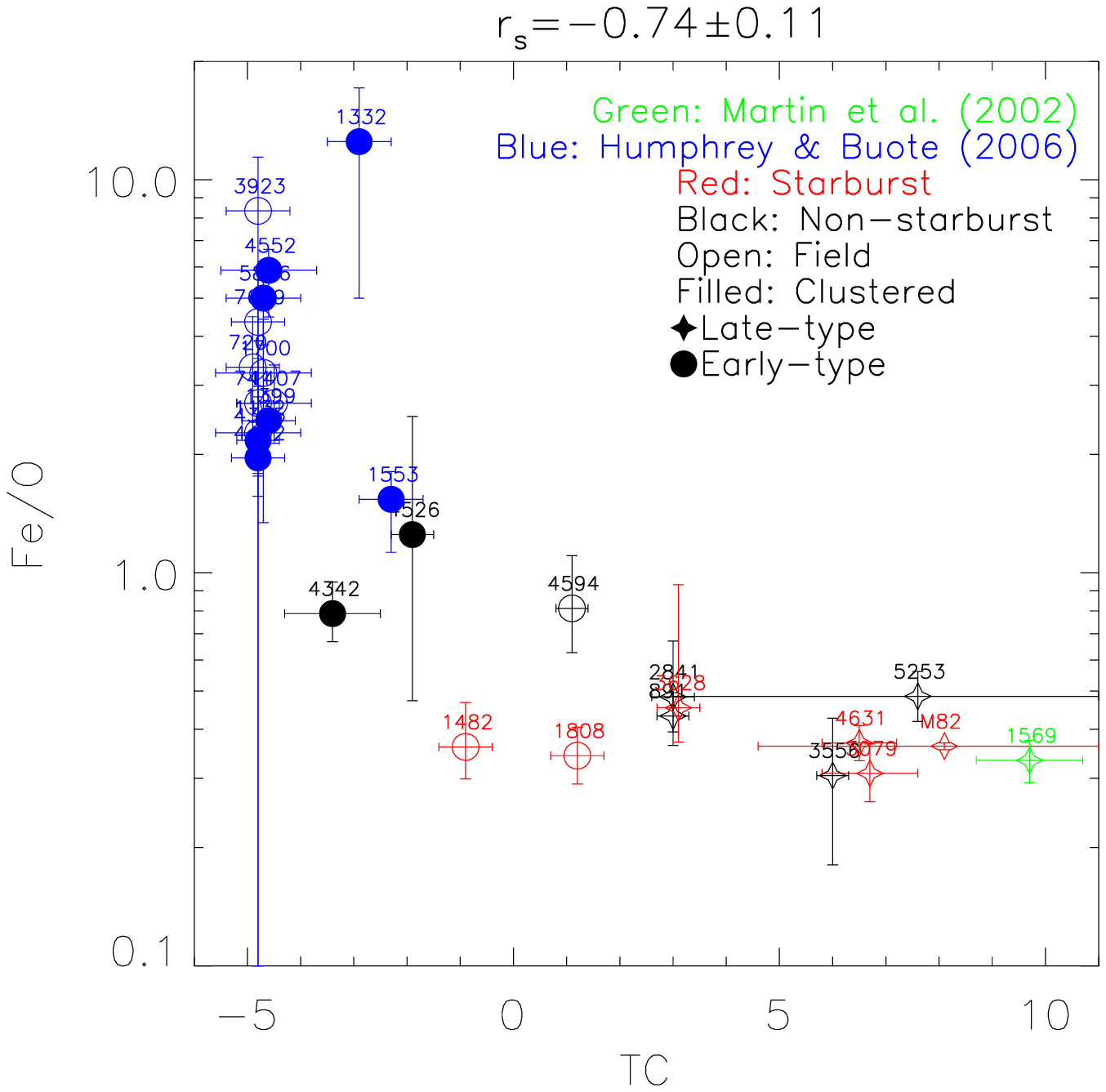,width=0.55\textwidth,angle=0, clip=}
\caption{Fe/O abundance ratio vs. morphological type code (TC) for selected galaxies (see text for details).}\label{fig:AbundanceRatio}
\end{center}
\end{figure}

Fig.~\ref{fig:AbundanceRatio} plots the Fe/O abundance ratios obtained for 14 galaxies with relatively high-quality X-ray spectra (Paper~I; NGC~5866 is not plotted here because of the large uncertainty in its Fe/O ratio) against the galaxy morphological type code, TC. \emph{Chandra} measurements of the abundance ratios for several other galaxies are also included \citep{Martin02,Humphrey06}, in order to expand the range of galaxy type for comparison. Note that \citet{Grimes05} also obtained the $\alpha/Fe$ abundance ratio for their sample galaxies, but the morphological classification is quite uncertain for many of their galaxies which are significantly distorted, so they are not included in Fig.~\ref{fig:AbundanceRatio}. All their sample galaxies have high $\alpha/Fe$ ratios (typically $\gtrsim2.5$), or low Fe/O ratios (typically $\lesssim0.4$). Fig.~\ref{fig:AbundanceRatio} shows a significant anti-correlation between Fe/O and TC ($r_s=-0.74\pm0.11$): i.e., the earlier the galaxy morphology, the higher the Fe/O ratio. The ratios of starburst galaxies are nearly constant ($\sim0.36\pm0.12\rm~solar$), consistent with \citet{Grimes05}'s measurements of their starburst galaxies.

\section{Discussion}\label{PaperIIsec:Discussion}

\subsection{Comparison to Elliptical Galaxies}\label{PaperIIsubsec:DiscussionElliptical}

Elliptical galaxies are often thought to have formation histories and/or environments substantially different from disc galaxies. Their coronae may also have distinguishable X-ray properties (e.g., \citealt{Mathews03}). It can thus be instructive to compare our measurements with those of elliptical galaxies \citep{OSullivan03,Boroson11}.

\begin{figure}[!h]
\begin{center}
\epsfig{figure=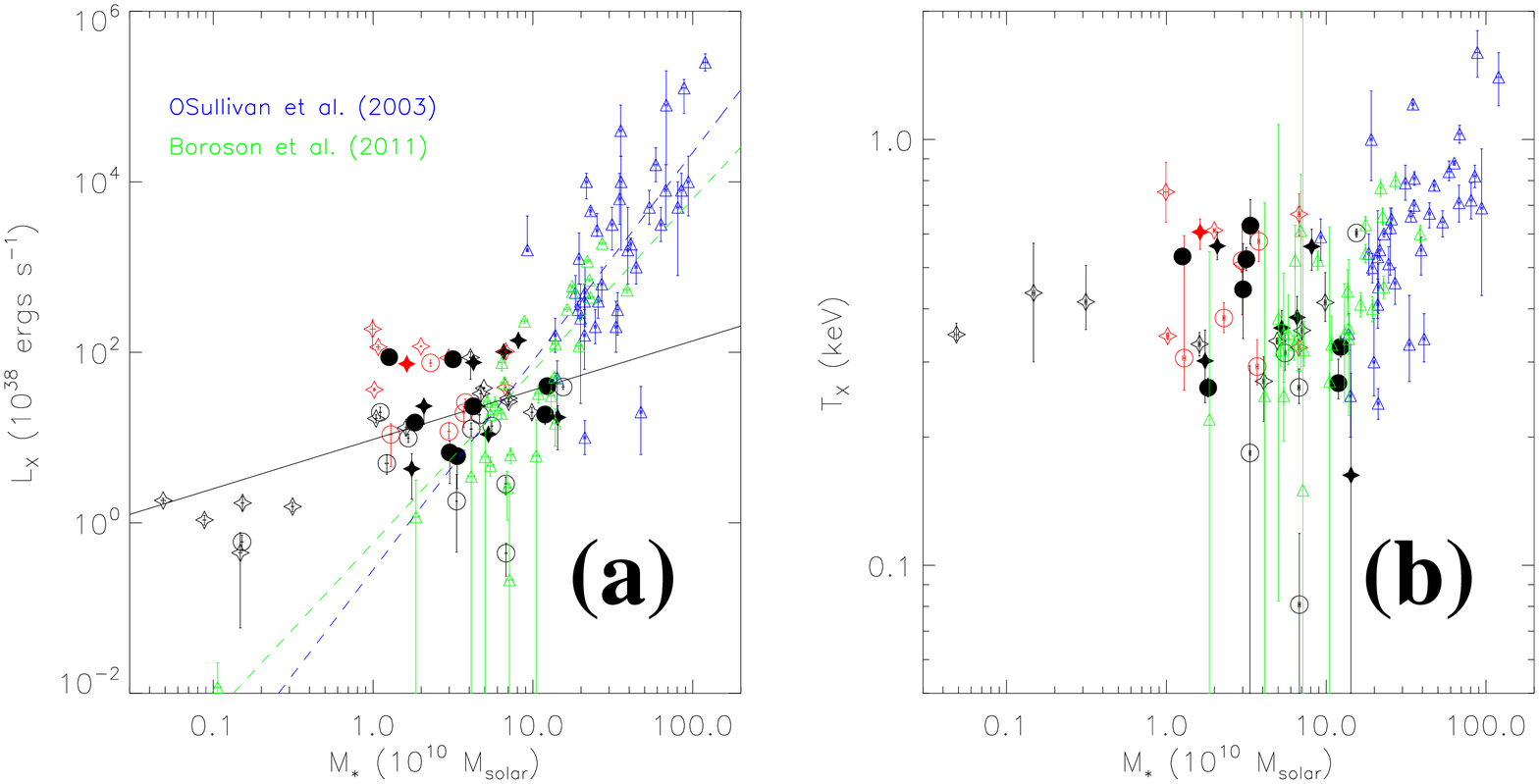,width=1.0\textwidth,angle=0, clip=}
\caption{X-ray luminosities (a) and temperatures (b) of the coronae vs. the stellar mass of the galaxies. Samples of elliptical galaxies \citep{OSullivan03,Boroson11} are plotted for comparison. The black solid line and the colored dashed lines in (a) represent the best-fit to the present, \citet{OSullivan03}, and \citet{Boroson11} samples, respectively. The K-band luminosity (used to calculate the stellar mass, Paper~I) of \citet{Boroson11}'s sample is obtained from the original paper, while that of \citet{OSullivan03}'s sample is obtained from the \emph{2MASS} extended source catalogue \citep{Skrutskie06}.}\label{fig:elliptical}
\end{center}
\end{figure}

The hot gas in and around massive elliptical galaxies is typically in hydrostatic equilibrium, unless disturbed by strong AGN feedback or galaxy merger (e.g., \citealt{Forman05,Kraft05}). They are therefore expected to have well-defined scaling relations between the coronal properties ($T_X$ or $L_X$) and galaxy mass. Such relations are indeed confirmed for X-ray luminous elliptical galaxies, as well as groups/clusters of galaxies (e.g., \citealt{OSullivan03,Sanderson03a}). In comparison, coronae of disc galaxies show a more complicated behavior. As shown in Fig.~\ref{fig:elliptical}, there is a significant morphological dependence of the coronal properties. In particular, disc galaxies, even if only non-starburst ones in the field, tend to be more X-ray luminous than elliptical galaxies of similar stellar masses in the $M_*\lesssim10^{11}\rm~M_\odot$ range (Fig.~\ref{fig:elliptical}a). But for more massive galaxies, this trend reverses. Quantitatively, the slope of the $L_X-M_*$ relation for disc galaxies ($0.58\pm0.05$) is significantly flatter than those for elliptical ones [$2.4\pm0.3$ for \citet{OSullivan03}'s sample and $2.0\pm0.2$ for \citet{Boroson11}'s sample]. Furthermore, massive and dwarf elliptical galaxies follow the same $L_X-M_*$ relation over almost three orders of magnitude in $M_*$. While the higher luminosities of elliptical galaxies in the $M_*\gtrsim10^{11}\rm~M_\odot$ range can be partially attributed to the inclusion of gas-rich cD-type galaxies \citep{OSullivan03}, the departure of disc galaxies from the scaling relation at low mass end probably indicates the important role that the SF and/or cool gas play in determining $L_X$ of a galaxy (see more discussions in \S\ref{PaperIIsubsec:DiscussionOrigin}).

As shown in Fig.~\ref{fig:elliptical}b, the $T_X-M_*$ correlation is absent for disc galaxies ($r_s=-0.20\pm0.16$), but is fairly strong for elliptical ones [$r_s=0.61\pm0.12$ for \citet{OSullivan03}'s sample and $0.52\pm0.16$ for \citet{Boroson11}'s sample]. Hot gas in disc galaxies tends to have a higher characteristic temperature than in elliptical ones of similar stellar masses. This clearly suggests the importance of additional heating sources other than Type~Ia SNe and stellar orbital motion (the typical gas heating mechanisms of low mass elliptical galaxies with virial temperatures $\lesssim10^6\rm~K$; \citealt{Mathews03,Bregman09b}), most likely massive stellar winds and CC SNe from young stellar objects.

\subsection{Fe/O Ratio across Galaxy Type}\label{PaperIIsubsec:DiscussionMetallicity}

As shown in Fig.~\ref{fig:AbundanceRatio}, the measured Fe/O ratios of the non-starburst galaxies show a significant correlation with the galaxy morphology types, while the ratios of starburst galaxies are nearly constant ($\sim0.36\pm0.12\rm~solar$). To our knowledge, this correlation has not been reported in existing publications. While the Fe/O ratios are often measured in various studies, they tend to focus on narrow ranges of galaxy types: e.g., ellipticals \citep{Humphrey06,Ji09} or starbursts \citep{Strickland04a,Grimes05}. Therefore, it is not surprising to have missed the $\rm Fe/O-TC$ correlation. The presence of the correlation may be attributed to several processes:

(1) \emph{Stellar feedback.} This includes ejecta of SNe, stellar winds from massive stars, and mass loss of evolved low-mass stars. While massive stars (primarily via CC SNe) are responsible for the bulk of the O-like elements, low-mass stars (via Ia~SNe) are for the Fe enrichment. These different metal enrichments from CC and Ia SNe can naturally produce the abundance pattern observed in the coronae of different types of galaxies (e.g., \citealt{Finoguenov99,Martin02,Kim04,Sato07}). With a typical lifetime of Type~Ia SNe progenitors, a canonical initial mass function (IMF; \emph{Salpeter} or \emph{Scalo}), and a usual SF history (an instantaneous burst, a constant SFR, or a more realistic SFR derived in the framework of chemical evolution models), the Fe/O ratio of the SN-enriched interstellar medium (ISM) is calculated to be $\gtrsim0.4\rm~solar$; the lower limit represents a negligible contribution from Ia~SNe \citep{Gibson97,Matteucci01}. This lower limit matches our measured value of ${\rm Fe/O}\sim0.36\rm~solar$ for the coronae of starburst galaxies and \citet{Grimes05}'s measurements of their starburst galaxy sample. For these galaxies, the chemical enrichment from massive stars is expected to dominate at the present. In addition, \citet{Matteucci01}'s models can also reproduce the correlation between the Fe/O ratio and the total Ia~SN contribution, as indicated by the Fe/O-morphology correlation (Fig.~\ref{fig:AbundanceRatio}).

(2) \emph{Mass loading of the cold ISM and destruction of dust grains.} In cold interstellar gas, iron is often highly depleted (as compared to oxygen) due to the formation of dust grains. The gas phase Fe/O ratio of the Galactic ISM, for example, is typically $<0.1\rm~solar$ \citep{Jenkins09}. Strong UV radiation in \ion{H}{2} regions can destruct dust grains and hence increase the Fe/O ratio to various degrees (e.g., ${\rm Fe/O}\sim0.34-2.51\rm~solar$ for a sample of Galactic \ion{H}{2} regions according to \citealt{Stanghellini10}; ${\rm Fe/O}\sim0.10-0.22\rm~solar$ for some extragalactic ones according to \citealt{Esteban02}). The mass loading of cool gas may then lead to a relatively broad range of the observed abundance ratios of the plasma in and around normal galaxies.

The mass-loading of dusty gas can also occur outside SF regions. Lines of observational evidence exist for the presence of extraplanar dust, at distances as far as 10~kpc away from actively SF galactic disc \citep{Howk97,Howk99,Heckman00} or even S0 galaxies \citep{Li09}. Such dusty cool gas may be eventually loaded into coronal gas via turbulent mixing and thermal conduction. The typical sputtering timescale is $\lesssim10^7\rm~yr$ for dust grains smaller than $\sim0.3\rm~\mu m$ after they become mixed microscopically with the coronal gas \citep{Popescu00}. This timescale is significantly shorter than the typical radiative cooling timescale of the gas ($\sim\rm Gyr$, Paper~I), but can be comparable to, or even longer than, the dynamic time of the galactic outflows or superwinds through a typical observed X-ray-emitting corona of a starburst galaxy \citep{Strickland00a}. Therefore, the metal abundance pattern in the coronal gas could range from being strongly Fe-depleted to reflecting a dust-free ISM.

In short, the present stellar feedback likely plays a key role in metal enrichment of the galactic coronae [but there is another possibility recently raised by \citet{Crain13} that the metallicity of galactic corona inferred from X-ray observations is nonproportionally affected by galactic feedbacks (than by accretion of external gas) because they are typically more effective in radiating X-ray]. In addition, additional feedback from young stellar populations also provides a consistent explanation for the higher luminosity and temperature of disc galaxies than those of elliptical ones (\S\ref{PaperIIsubsec:DiscussionElliptical}).

\subsection{X-ray Radiation Efficiency}\label{PaperIIsubsec:DiscussionScaling}

A low X-ray radiation efficiency (far below unity) is commonly seen in nearby galaxies (e.g., \citealt{Mineo12}) and known as the \emph{``missing galactic feedback''} problem, which is especially significant in galactic spheroids with little cool gas to consume or convert the feedback energy (see \citealt{Wang10} and references therein). The ``missing'' feedback is most likely carried out by hot, tenuous, and metal-enriched galactic outflows (e.g., \citealt{Strickland00a,Strickland07,Strickland09}). Under this scenario, we further explore the physical meaning of the dependence of $\eta$ on $M_{TF}/M_*$ and $F_{SN(CC)}$ (Fig.~\ref{fig:efficiency}d,f), or the joint scaling relations (Eqs.~\ref{equi:ESNMTFMstellarLX} and \ref{equi:ESNFSNCCLX}; \S\ref{PaperIIsubsec:BasicScaling}).

To understand the physical meaning of $M_{TF}/M_*$, we start with the Tully-Fisher relation: $M_{TF}\propto v_{rot}^\alpha$ \citep{Tully77}. Assuming the rotation velocity $v_{rot}$ is measured at a radius $r_{max}$, the gravitational mass within $r_{max}$ can be expressed as: $M_g\propto v_{rot}^2r_{max}$. On the other hand, the stellar mass is measured from the K-band luminosity ($L_K$) with a mass-to-light ratio of $c_K$, so $M_*\propto c_KI_Kr_K^2$, where $I_K$ is the mean K-band surface brightness within the aperture with radius $r_K$. We could then derive the following dependence:
\begin{equation}\label{equi:TullyFisher}
M_{TF}/M_*\propto (1/c_K)(r_K/r_{max})^2I_Kv_{rot}^{\alpha-4}(M_g/L_K)^2.
\end{equation}
In this paper, we have adopted a color dependent mass-to-light ratio, so $c_K$ is not constant, but has a typical scatter of only a factor of $\lesssim2$ for our sample galaxies \citep{Bell01}. $r_K/r_{max}$ depends on the photometry aperture and the rotation curve, so is not expected to have any systematic bias. For the baryonic Tully-Fisher relation, $\alpha\sim3.51$ \citep{Bell01}, so $M_{TF}/M_*$ has only weak dependence on $v_{rot}$; we thus neglect this term in the following discussions. Finally, we only need to consider two terms in Eq.~\ref{equi:TullyFisher}: the K-band surface brightness $I_K$ and the mass-to-light ratio $M_g/L_K$. We herein speculate two possible explanations of the $\eta-M_{TF}/M_*$ relation. \emph{Firstly}, with a roughly constant $c_K$, $I_K$ is proportional to the mass density in the inner region of a galaxy (within the K-band photometry aperture). The $\eta-M_{TF}/M_*$ correlation thus indicates a gravitational confinement effect, i.e., the higher the central mass density, the slower the outflow and so the larger fraction of SN energy radiated. \emph{Alternatively}, the $\eta-M_{TF}/M_*$ correlation is related to $M_g/L_K$. In the inner region, the gravity is often dominated by stellar mass, so $M_g\thickapprox M_*$ and $M_g/L_K$ is roughly constant. But \citet{McGaugh00} found a significant departure of low mass gas-rich galaxies from the optical Tully-Fisher relation, which is explained by the need to include cool gas mass in the galaxy baryon content. Therefore, $M_g/L_K$, and so $M_{TF}/M_*$, is proportional to the richness of cool gas in a galaxy. The positive dependence of $\eta$ on $M_{TF}/M_*$ thus means that the X-ray emission is related to the presence of cool gas. As will be discussed in \S\ref{PaperIIsubsec:DiscussionOrigin}, cool-hot gas interaction can indeed significantly enhance the emission. This scenario is further supported by the improved $\eta-M_{TF}/M_*$ correlation for starburst galaxies (Fig.~\ref{fig:efficiency}e), which often have turbulent interstellar structures (e.g, NGC~520, NGC~660, refer to Paper~I for description of individual galaxies), and are thus expected to have increased cool/hot gas interfaces. Furthermore, we notice that early-type starburst galaxies all have a relatively low $\eta$ (Fig.~\ref{fig:efficiency}e), probably because of their relatively low cool gas content as compared to late-type ones.

The physical meaning of $F_{SN(CC)}$ is more apparent. By definition, it represents the driving force for the superbubble blowout from a galactic disc \citep{MacLow88,Strickland04b}. In such a blowout scenario, a superbubble is created by SNe and fast stellar winds of massive star clusters in the stratified atmosphere of a galactic disc. Blowout occurs if the shell of the swept-up material begins to accelerate down the density gradient. At this point the Rayleigh-Taylor instability sets in and fragments the shell. The hot gas in the bubble then vents out into the surrounding low-density halos. Further modification to this scenario includes the cooperation among multiple star clusters in driving a merged single supergiant bubble, which enables non-nuclear-starburst galaxies with disc-wide SF (while individual star clusters of which may not be energetic enough) to blow out hot gas. Therefore, there may exist a critical $F_{SN}$ value ($F_{SN,crit}$) above which such a bubble can blow out. This blowout threshold is only affected by the density and thermal pressure (temperature) of the ambient medium \citep{Strickland04b}. The anti-correlation between $\eta$ and $F_{SN}$ (Fig.~\ref{fig:efficiency}e), although weak, is thus quite suggestive, and can be naturally explained with the above SF-driven superbubble blowout scenario. In other words, an increase of the energy injection rate per unit area tends to raise the specific energy or temperature of the blown-out gas, leading to a faster expansion and hence a reduction of the X-ray emissivity. A similar scenario is often suggested for AGN feedback to reduce the cooling flow in the ICM (e.g., \citealt{McNamara12} and references therein). \citet{Boroson11} also find an age dependence of the coronal luminosity of their median-mass subsample of elliptical galaxies. The lower coronal luminosities of younger galaxies are thought to be due to the emptying of the ISM by recent SF feedback.

Before going on to discuss other potential processes that may further affect the coronal X-ray emission, we need to consider some possible systematical biases which may affect the dependence of $\eta$ on $M_{TF}/M_*$ and $F_{SN(CC)}$.

In \S2.2.2 of Paper~I, we have discussed uncertainties in the measurement of $v_{rot}$ and thus $M_{TF}$, which could affect the $\eta-M_{TF}/M_*$ correlation. Here we further consider the uncertainties in the \emph{SF size} estimation, which directly affects the measurement of $F_{SN}$ and hence the $\eta-F_{SN}$ correlation. In the present study, we have estimated $F_{SN}$ using the area of a galactic disc, i.e., $D_{25}$ from the optical measurement. But in the disc blowout scenario, it should be the active SF area involved in the generation of supergiant bubbles \citep{Strickland04b}. The estimation of the SF area is not straightforward, however, because the edge-on perspective results in a heavy extinction and projection confusion, and because individual SF regions are often not well resolved even in nearby galaxies. \citet{Strickland04b} tried to use the radio extension to define the SF region, but obtained a critical $F_{SN}$ largely \emph{inconsistent} with their analytical calculation. In contrast, the adoption of $D_{25}$, uniformly measured, provided a fairly good match (the two red solid lines in Fig.~\ref{fig:efficiency}f). This is not unexpected because the blowout scenario of the galactic coronae is a result of the cooperation of star clusters, which could be spread widely over a galactic disc. For ease of comparison for our entire sample, we have also adopted $D_{25}$ as the SF size estimate. For a careful test of the blowout scenario, especially for individual galaxies, one would need to get a better size estimate (e.g., based on high spatial resolution mid-IR imaging observations).

Our measured X-ray radiation efficiency differs by a factor of $\sim10$ from recent results ($\sim5\%$) obtained by \citet{Mineo12} from a \emph{Chandra} study of 29 star forming galaxies with various inclination angles. Our measurements do not include a highly attenuated component in galactic discs, especially in immediate vicinities of recent SF regions, where X-ray absorption could be substantially higher than what we have assumed for the coronae or inferred from the fits to the global spectra \citep{Mineo12}. This component is better examined with galaxies of low inclination angles (e.g., \citealt{Owen09}). There are also other differences in these two studies: the inclusion of Type~Ia SNe in our analysis, the adoption of the bolometric luminosity (which can be very sensitive to the spectral model selection and the absorption correction) in \citet{Mineo12} instead of our adopted 0.5-2~keV luminosity (which is expected to be a factor of $\sim1.4-3$ times lower), the galactic disc inclinations, and the adopted photometry apertures.

\subsection{Origins of Enhanced X-ray Emission}\label{PaperIIsubsec:DiscussionOrigin}

We have noted unusual multi-wavelength characteristics of individual galaxies in \S4 of Paper~I. Some of these characteristics may cause substantial departures of the galaxies from the joint scaling relations (e.g., Eq.~\ref{equi:ESNMTFMstellarLX}). Take NGC~4438 as an example, which shows the largest departure from the relation (Fig.~\ref{fig:2Dfit}). As discussed in Paper~I, this galaxy is a member of the Virgo cluster and is interacting with its companion NGC~4435 \citep{Machacek04}. The tidally stripped cool gas is apparently undergoing strong interaction with the ICM (as revealed by the large-scale X-ray tail), probably through mass loading or CXE. Such processes may be responsible for the enhanced X-ray emission, which is substantially stronger than expected from the galaxy's moderate SFR and mass. Similar effects (in addition to the SN feedback and gravitational confinement considered above) may also contribute to the X-ray emission from other galaxies. We herein examine these processes in some details:

(1) \emph{Mass loading.} Stellar feedback is often accompanied with additional mass input (in addition to SN ejecta); e.g., via shock heating of the surrounding ISM. Mass-loading to hot gas can also occur at its interfaces with cool gas via thermal conduction and turbulent mixing (e.g., \citealt{Melioli04}). Such mass loading can significantly enhance the X-ray emission \citep{Li11}, especially for gas-rich galaxies (e.g., NGC~5866, \citealt{Li09}; NGC~4388, \citealt{Machacek04}; also see Paper~I). This may help to explain the higher X-ray luminosity of disc galaxies than elliptical ones in the same stellar mass range (\S\ref{PaperIIsubsec:DiscussionElliptical}). If the observed X-ray luminosity represents the thermal emission, the mass-loading may also be needed to explain the measured range of coronal temperature, which is substantially lower than that expected from the SN feedback \citep{Strickland00a}. But physically, the mass-loading alone may be difficult to realize the narrow range of the temperature.

(2) \emph{Charge exchange.} We have assumed that the X-ray emission from the coronae arises from collisionally excited (thermal) plasma. But this assumption may be problematic if the contribution from the CXE is important. In addition to increasing the soft X-ray luminosity (\S5.3 of Paper~I), the CXE could also affect the measurement of the thermal and chemical states of the coronal gas by greatly enhancing the inner-shell transitions of O, and possibly Ne and Mg \citep{Liu10,Liu11,Liu12}. Active SF galaxies with typically large cool gas contents tend to have a large contribution from the CXE, and show spatial correlation between soft X-ray intensity and cool gas, which can be traced by H$\alpha$ or dust emission \citep{Strickland00b,Strickland02,Strickland04a}. The transitions after the CXE occur in lower ionization states than those in the thermal emission (because of the captures of electrons by the thermal ions). Therefore a significant CXE contribution tends to result in an underestimate of the temperature of the hot gas, if its X-ray spectrum is fitted with a thermal plasma model. Therefore, the true temperatures of the coronae can be significantly higher than those measured in Paper~I, which makes the discrepancy between the disc galaxies and the elliptical ones in the $L_X-T_X$ plot (Fig.~\ref{fig:TX}d) even more profound.

The CXE can also affect the metallicity measurement by significantly enhancing the \ion{O}{7} forbidden and inter-combination lines (e.g., \citealt{Liu10,Liu11,Liu12,Konami11}). The nearly constant Fe/O ratio measured for the coronae of starburst galaxies may be partly a result of the saturated CXE contribution in case of active SF and large amount of cool gas. If the soft X-ray emission primarily arises from the interfaces instead of the hot plasma itself, the metal abundance cannot be measured by assuming a simple thermal model. If we assume the intrinsic Fe/O ratio to be $\sim$solar (as expected in the ISM with little dust depletion), for example, the constant Fe/O ratio of $\sim0.36\rm~solar$ then implies that $\sim60\%$ of the oxygen line emission may be produced by the CXE, consistent with the estimate based on high-resolution X-ray grating spectra of M82 \citep{Liu11}.

(3) \emph{Environmental effects.} The X-ray emissivity enhancement of a corona can be due to the confinement by the thermal/ram-pressure of the ICM \citep{Mulchaey10,Lu11}, the presence of cool gas surrounding a galaxy (e.g., M82; \citealt{Lehnert99}), or the accretion of external gas \citep{Bogdan11}. It is suggested that the offset in the $L_X-T_X$ relations separately defined by X-ray luminous \citep{OSullivan03} and faint \citep{Boroson11} elliptical galaxies (Fig.~\ref{fig:TX}d) cannot be attributed only to the better removal of stellar sources in the latter, but may also be partly intrinsic because many galaxies in the former sample are in dense cluster environments. On the other hand, strong ram-pressure stripping can also efficiently remove the coronal gas. Both ram-pressure stripping and thermal confinement are proportional to the mass of a cluster (ram-pressure is proportional to the velocity square of the galaxy's orbital motion in the ICM, while thermal confinement is proportional to the temperature of the ICM); their relative importance is therefore not affected by the host galaxy halo mass. \citet{Bahe12a,Bahe12b} have shown that the ram-pressure dominates the confinement pressure for most of the satellite galaxies in orbit about galaxy groups and clusters, and even the $\sim16\%$ galaxies in reverse situations are typically X-ray faint because they have already experienced strong ram-pressure stripping during early orbital motion. Furthermore, the disturbances due to tidal forces of companion galaxies may also affect the dynamics of the coronal gas, as well as the stability of the cool gas disc, thus the efficiency of SF and its feedback. Given all these complications, it may be expected that we find no significant correlation between $L_X$ and the environmental density $\rho$ ($r_s=0.31\pm0.15$ for the whole sample), indicating that either the environmental effects have a large diversity (both positive and negative roles on $L_X$) or that their effects on $L_X$ are relatively minor. We will further discuss the role of environmental effects in Paper~III by comparing our sample with theoretical predictions of X-ray properties of field disc galaxies from hydrodynamical simulations.

\section{Summary and Conclusions}\label{PaperIIsec:Summary}

Based on our \emph{Chandra} measurements of diffuse X-ray emission from 53 nearby highly-inclined disc galaxies (Paper~I), we have conducted a correlation analysis of their coronal and other multi-wavelength properties. Comparisons are also given with corresponding properties of elliptical galaxies. We have finally explored the implications of our findings and have further discussed various uncertainties. Our main results and conclusions are summarized below:

(1) A strong correlation is present between the coronal soft X-ray luminosity ($L_X$) and the SFR. The same linear relation holds for both starburst and non-starburst galaxies. In contrast, the correlations of $L_X$ with the galaxy masses ($M_*$ or $M_{TF}$) are weak for the entire sample. But if only non-starburst galaxies are considered, these three correlations are then comparable. Starburst galaxies show significant deviations (offset to high $L_X$) from the $L_X$-mass correlations.

(2) A tighter correlation is found between $L_X$ and $\dot{E}_{SN}$ (than between $L_X$ and SFR), the total SNe mechanical energy input rate including contributions from both core collapsed and Type~Ia SNe.

(3) The X-ray radiation efficiency ($\eta\equiv L_{X}/\dot{E}_{SN}$) is $\sim0.4\%$ with an $rms$ of $0.50\pm0.06~\rm dex$. $\eta$ shows little correlation with SFR, $M_*$, or $M_{TF}$, but does correlate with $M_{TF}/M_*$ and with the core collapsed SN surface rate ($F_{SN(CC)}$), which can be characterized as: $\eta=(0.41_{-0.12}^{+0.13}\%)M_{TF}/M_*$ and $\eta=(1.4\pm0.5\%)F_{SN(CC)}^{-(0.29\pm0.11)}$.

(4) Accounting for all the above dependences, $L_X$ of disc galaxies can typically be predicted with rather small uncertainties. In particular, the starburst subsample shows a surprisingly tight ($rms=0.32\pm0.06\rm~dex$) scaling relation of $L_X=(0.41_{-0.12}^{+0.13}\%)\dot{E}_{SN}M_{TF}/M_*$ (with a correlation coefficient $r_s=0.68\pm0.16$ compared to $-0.03\pm0.30$ for the $L_X-{\rm SFR}$ relation).

(5) The X-ray spectrum of a corona can typically be characterized by an optically-thin thermal plasma. The inferred characteristic temperature ($T_X$) shows little dependence on the total/specific SFR, $L_X$, or cold gas mass of a galaxy.

(6) The Fe/O abundance ratio shows a dependence on the morphological type for non-starburst galaxies: i.e., the earlier a galaxy, the Fe-richer its corona. Furthermore, starburst galaxies have a nearly constant Fe/O ratio of $\sim0.36\pm0.12\rm~solar$, which is considerably lower than that measured in non-starburst galaxies. Both the abundance-morphology correlation and the low Fe/O ratio of starburst galaxies could be reasonably explained with the expected different metal enrichments from CC and Ia~SNe.

(7) The coronae of disc galaxies tend to be more X-ray luminous and Fe-rich and have higher characteristic temperatures than those of elliptical ones of similar masses in the galaxy mass range of $M_*\lesssim10^{11}\rm~M_\odot$.

(8) Additional processes, such as mass loading and CXE at the interfaces between hot and cool gases, as well as various environmental effects, may be effective in regulating the observed X-ray properties of some galactic coronae. In particular, the interface effects tend to result in a low characteristic temperature and Fe/O abundance ratio (when inferred from the X-ray spectral fits), as well as the enhancement of X-ray emission from disc galaxies as compared to elliptical ones. While environmental effects may strongly affect some individual galaxies (e.g., NGC~4438), we do not find a significant correlation between $L_X$ and the environmental density ($\rho$). Apparently, $\rho$ has both positive and negative effects on $L_X$, or the environmental effects are generally minor for the whole sample.

In this paper, we have examined the properties of galactic coronae in comparison with stellar feedback in disc galaxies. In particular, we have introduced the parameter $\eta$, based on a primarily linear correlation between $L_X$ and $\dot{E}_{SN}$. The luminosities, as well as the thermal and chemical properties of the coronae, can be self-consistently explained with a feedback-dominated scenario, partly because we are only exploring the coronae on relatively small scales near galactic disks [the coronae have a typical vertical extent (expressed in five times the exponential scale height of the vertical brightness profile) $\sim1-30\rm~kpc$ with a median value of $\sim5\rm~kpc$, Paper~I] and also because very few massive (e.g., $M_*\gtrsim10^{11}\rm~M_\odot$) disc galaxies have been carefully studied, which are rare. But we do find some hints for the importance of gravitational effects, e.g., the correlation between $\eta$ and $M_{TF}/M_*$ as explained in the gravitational confinement scenario. In a follow-up paper (Paper~III), we intend to present comparisons of our X-ray measurements with galaxy formation models and with cosmological numerical simulations, as well as with recent claimed detection of large coronae around several massive disc galaxies.

\acknowledgements

We thank Robert Crain, Ying Zu, Ian McCarthy, and Yannick Bah$\rm\acute{e}$ for many helpful discussions, as well as the referees of both the present paper and Paper~I for their constructive comments and suggestions. This work is supported by NASA through the CXC/SAO grant AR0-11011B, and the ADAP grant NNX12AE78G.

\end{document}